\documentclass[showpacs,eqsecnum,prb]{revtex4}
\usepackage[dvips]{graphicx}
\usepackage{amsmath,amssymb}
\begin{document}
\title{Photon energy absorption rate of a striped Hall gas}
\author{Y. Ishizuka,$ ^{1}$ T. Aoyama,$ ^{2}$ N. Maeda$ ^{1}$ 
and K. Ishikawa$ ^{1}$}

\affiliation{ $ ^{1}$Department of Physics, Hokkaido University, 
             Sapporo 060-0810, Japan\\
 $ ^{2}$Institute Henri Poincar\'e, 11  rue Pierre et Marie Curie,
75231 Paris cedex 05 France }

\date{\today}

\begin{abstract}
Using symmetries of
the current correlation function, we analyze the frequency dependence
 of the photon energy absorption rate of a striped Hall gas.
Since the magnetic translational symmetry is 
spontaneously broken  in the striped Hall gas,
 a Nambu-Goldstone (NG) mode appears. 
It is shown that the NG mode causes a sharp absorption 
 at the zero energy in the long wavelength limit 
by using the single mode approximation.
The photon energy absorption rate at the NG mode frequency
strongly depends on the direction of the wave number vector.
Whereas, the absorption rate at the cyclotron frequency
does not depend on the direction of the wave number vector 
in the long wavelength limit.  
The cyclotron resonance is not affected in the striped Hall gas.
Our result supplements the Kohn's theorem in the system of the NG mode.
\end{abstract}
\maketitle

\section{Introduction}
The two-dimensional electron system under a strong magnetic field, 
the quantum Hall system (QHS), without impurities has four symmetries,
that is
the electromagnetic $U(1)$, 
the magnetic rotation, and the magnetic translations 
in $x$ and $y$ direction. 
The magnetic translational symmetry in one direction and the magnetic
rotational symmetry are spontaneously broken  in 
a striped Hall gas,  which is supposed to be realized at the
 half-filling of the third and higher Landau Level (LL)s. 
The striped Hall gas has an anisotropic Fermi surface, and leads to
 a highly anisotropic resistivity.~\cite{Stripe1,Stripe2}  
Since the magnetic translation 
is spontaneously broken  in this state,
a Nambu-Goldstone (NG) mode  appears. 
Dynamical effects of  the NG mode are studied in the present paper.

There are several theoretical works which would  lead to the 
highly anisotropic resistivity at
the half-filled third and higher LLs. The Hartree-Fock
approximation~(HFA) at the half-filled $l$th LL predicts two solutions,
a unidirectional charge density
wave~(UCDW)~\cite{HFref1,HFref2,HFref3,HFref4} and a highly anisotropic
charge density wave~(ACDW).~\cite{phonon2}  The striped Hall gas is the
UCDW which has an energy gap in one direction and no energy gap in
another direction. The ACDW has an energy gap in each direction.
Collective modes for the UCDW have been studied based on the edge current
picture,~\cite{phonon1,rev,phonon3,phonon4,phonon5} 
the generalized random phase approximation~(GRPA),~\cite{phonon7}  and
 the single mode approximation~(SMA).~\cite{SMA,SMA1,SMA2}
Collective modes for the ACDW have been studied in the time-dependent
HFA.~\cite{phonon2}
 We use  the UCDW and the SMA 
 to study the photon absorption effect of the NG mode.

 The Kohn's theorem~\cite{KOHN}
 is a quite general theorem concerning the photon absorption in the QHS.
    It includes two statements on the cyclotron resonance.  
    The first one is that a sharp   absorption of a homogeneous
    rotating microwave occurs only at the cyclotron 
    frequency. The second one is that the cyclotron resonance is not
    affected by electron interactions.
The implicit assumption for the first statement is that 
the zero-energy excited state is absent in the long wavelength limit.
The assumption is not satisfied in the striped Hall gas.

It is an open question if  
 a sharp absorption at the NG 
mode frequency occurs in the long wavelength limit. 
A photon energy absorption rate is proportional to the  imaginary part
of a current correlation
function, and is inversely proportional to the frequency~$\omega$ of the 
electromagnetic vector potential 
${\bf A}_{ext}={\mathcal E} e^{i{\bf k}\cdot {\bf r}+ik_{z}z-i\omega
t}/i\omega$. Here ${\mathcal E}$ is the polarization vector,
 $({\bf k},k_{z})$  is the wave number vector
  and $({\bf r},z,t)$  is the space-time coordinate.
We project the system onto the $l$th LL and study the contribution
of the NG mode.
Since the imaginary part of the LL projected current correlation
 function becomes zero in the $|{\bf k}|\rightarrow 0$ limit,
it would appear that the photon energy absorption rate is zero.
However, when $\omega$ equals  the NG mode frequency $\omega_{NG}({\bf k})$, 
the denominator also becomes zero in the $|{\bf k}|\rightarrow 0$ limit.
Hence, the $|{\bf k}|\rightarrow 0$ limit of 
the photon energy absorption rate could be finite.
Actually, 
we find that a sharp absorption occurs at the NG mode frequency. 
The photon energy absorption rate at the NG mode frequency
strongly depends on the direction of ${\bf k}$.

Without the LL projection, we study the cyclotron resonance in the 
striped Hall gas.
The photon energy absorption rate at the cyclotron frequency
does not depend on the direction of ${\bf k}$
in the $|{\bf k}|\rightarrow 0$ limit.  
Then we find  that the
cyclotron resonance  is unaffected in the striped Hall gas phase.
The second statement of the  Kohn's theorem is intact. 

This paper is organized as follows. 
In Sec.~II, the symmetries of the QHS are clarified 
and the striped Hall gas  is constructed as an eigenstate of charges
of unbroken symmetries. The current correlation
function is computed by means of the magnetic translational invariance of
the striped Hall gas in Sec.~III.  A trigonometric factor of the
momentum and the  periodic Dirac's delta function of wave number vectors 
 appear due to the stripe periodicity.
In order to investigate the current correlation function in the long 
wavelength limit, we also use the Hartree-Fock (HF) solution of the
striped Hall gas    in Sec.~III.
We discuss the photon energy absorption  rate of the striped Hall gas 
and the f-sum rule by
using the SMA spectrum in Sec.~IV and give a discussion and summary in Sec.~V.
Appendix A gives a definition of the photon energy absorption rate. 
The cyclotron resonance  is derived in appendix B.

\section{Broken symmetries of the striped Hall gas}
The striped Hall gas spontaneously breaks the magnetic translational
symmetry  of the QHS. We clarify symmetries of the QHS, 
and define the striped Hall gas state
in terms of conserved charges of the magnetic translational symmetry.

\subsection{Symmetries of the QHS}
In the QHS, the
Hamiltonian has four symmetries, that is the electromagnetic $U(1)$ symmetry, 
the magnetic translational symmetries in $x$ and $y$ direction, 
and  the magnetic rotational symmetry.   
First, we introduce conserved currents and charges 
for these symmetries.
The two-dimensional vector potential $A_{i}({\bf r})$ is introduced
by means of  $B=\partial_{x}A_{y}-\partial_{y}A_{x}$. We ignore the
spin degree of freedom  and use the natural unit $(\hbar=c=1)$ in the
present paper. We introduce  relative coordinates and
guiding-center coordinates of the electron cyclotron motion.
 The relative coordinates are defined by
$\xi =(-i\partial _{y}+eA_{y})/eB$ and 
    $\eta = -(-i\partial _{x}+eA_{x})/eB$.
The guiding-center coordinates are defined by
      $ X=x-\xi$ and
      $Y=y-\eta$.
 These coordinates satisfy the following commutation relations,
$[X,Y]=-[\xi,\eta]=i/eB$ and
$[X,\xi]=[X,\eta]=[Y,\xi]=[Y,\eta]=0$.
 The operators $X$ and $Y$ are the generators of the magnetic
translations of the one-electron state in $-y$ direction and 
$x$ direction, respectively. 
The total Hamiltonian $H$ of the QHS is the
sum of the free Hamiltonian $H_{0}$ and the Coulomb interaction
Hamiltonian $H_{int}$ as follows,
  \begin{eqnarray}
   H&=&H_{0}+H_{int}, \nonumber\\
   H_{0}&=&\int d^{2}r \Psi^{\dag}({\bf r})
            \frac{m\omega_{c}^{2}}{2}(\xi^{2}+\eta^{2})
            \Psi({\bf r}), \nonumber\\
   H_{int}&=&\frac{1}{2}\int d^{2}r
                         d^{2}r^{\prime}
             \Psi^{\dag}({\bf r})
             \Psi^{\dag}({\bf r}^{\prime})
             V({\bf r}-{\bf r}^{\prime})
             \Psi({\bf r}^{\prime})\Psi({\bf r}), 
  \end{eqnarray}
where $\Psi({\bf r})$ is the electron field operator, 
$\omega_{c}=eB/m$, and
$V({\bf r})=q^{2}/r$\  $(q^{2}=e^{2}/4\pi\epsilon$, $\epsilon$
 is the background dielectric constant).
We do not consider  impurities and a confining potential in this paper.
Noether currents of the electromagnetic  $U(1)$, the  magnetic
translations  in 
$x$ and $-y$ directions are defined by
 $ j^{\mu}=\Psi^{\dag}v^{\mu}\Psi$,
 $ j^{\mu}_{Y}=\Psi^{\dag}v^{\mu}Y\Psi
             +\delta^{\mu}_{x}{\cal L}/eB$ and
 $ j^{\mu}_{X}=\Psi^{\dag}v^{\mu}X \Psi
              -\delta^{\mu}_{y}{\cal L}/eB$.
Here,
$v^{\mu}=(1,{\bf v}),{\bf v}=\omega_{c}(-\eta,\xi)$
and ${\cal L}$ is the Lagrangian density for the total Hamiltonian $H$.
Conserved charges $Q,Q_{Y},Q_{X}$ for the electromagnetic $U(1)$ and
 the magnetic translations are defined respectively by
  $ Q=\int d^{2}r j^{0}({\bf r})$,
  $Q_{Y}=\int d^{2}r j^{0}_{Y}({\bf r})$  and
  $Q_{X}=\int d^{2}r j^{0}_{X}({\bf r})$.
These charges commute with the total Hamiltonian $H$ and obey
$[Q_{X},Q_{Y}]=iQ/eB$. The electromagnetic $U(1)$ charge $Q$ commutes with
all conserved charges. We assume that $Q$ is not broken and the
ground state
 is the eigenstate of $Q$ as $Q|0\rangle =N_{e}|0\rangle$, where
$N_{e}$ is the number of electrons.

\subsection{Symmetries of  the striped Hall gas}
The conserved charges of  magnetic translations satisfy the following
algebra,
 \begin{eqnarray}
  \label{eqn:ssb1}
   \frac{\partial}{\partial x}j^{\mu}({\bf r})
    &=&-\frac{2\pi}{i a^{2}}[j^{\mu}({\bf r}),Q_{Y}],\nonumber\\
   \frac{\partial}{\partial y} j^{\mu}({\bf r})
   &=&\frac{2\pi}{i a^{2}}[j^{\mu}({\bf r}),Q_{X}],
  \end{eqnarray}
where $a=\sqrt{2\pi/eB}$.
Those states which are periodic in one direction and uniform in another
direction break the  magnetic translational symmetry.
Since an expectation value of the left hand side of  
Eq.~(\ref{eqn:ssb1}) with respect to the states becomes nonzero,
the states cannot be eigenstates of the conserved charge $Q_{X}$ or
$Q_{Y}$.
Therefore the magnetic translational symmetry is spontaneously broken.

We construct the striped Hall gas state which breaks
the magnetic translational symmetry in $x$ direction and
 conserves the magnetic translational symmetry in $y$ direction. 
The magnetic translations in $x$ and $y$ directions
 are generated  respectively by 
$\exp{(i\frac{2\pi}{a^{2}}xQ_{Y})}$
 and 
$\exp{(-i\frac{2\pi}{a^{2}}yQ_{X})}$
which is derived from  Eq.~(\ref{eqn:ssb1}). 
Hence, the striped Hall gas state  is an  eigenstate of an
infinitesimal magnetic translation in $y$ direction,
$\exp{(-i\frac{2\pi}{a^{2}}yQ_{X})}$, and is also an eigenstate of
a stripe period  magnetic translation in $x$ direction, 
$\exp{(i\frac{2\pi}{a^{2}}r_{s}a n_{x}Q_{Y})}$, where $n_{x}$ is an
integer and $r_{s}a$ is a  period of the stripe.
The striped  Hall gas state is characterized by an
 eigenvalue ${\bf K}^{(m)}$ of
the magnetic translation generator 
and  a species index $\sigma$ as
  \begin{eqnarray}
  \label{eqn:states}
    e^{i\frac{2\pi}{a^{2}}r_{s}an_{x}Q_{Y}}
                 |{\bf K}^{(m)},\sigma\rangle
  &=& e^{ir_{s}an_{x}K_{x}^{(m)}}
                 |{\bf K}^{(m)},\sigma\rangle ,
  \nonumber\\
    e^{-i\frac{2\pi}{a^{2}}yQ_{X}}
                 |{\bf K}^{(m)},\sigma\rangle
  &=&e^{-iy{K}_{y}^{(m)}}
                 |{\bf K}^{(m)},\sigma\rangle ,
  \nonumber\\
     H|{\bf K}^{(m)},\sigma\rangle
  &=& E_{\sigma}({\bf K}^{(m)})
                     |{\bf Q}_{X}^{(m)},\sigma\rangle.
  \end{eqnarray}
The ground state obeys $ H|{\bf 0},\sigma_{0}\rangle
  =E_{\sigma_{0}}({\bf 0}) |{\bf 0},\sigma_{0}\rangle$.
We assume a periodic boundary condition for the striped Hall gas
 in a rectangle  with the length
$L_{x}ar_{s}$ and $L_{y}a/r_{s}$ as 
  \begin{eqnarray}
  e^{i\frac{2\pi}{a^{2}}Q_{Y}L_{x}r_{s}a}
                    |{\bf K}^{(m)},\sigma \rangle
  &=&|{\bf K}^{(m)},\sigma \rangle,
   \nonumber\\ 
   e^{-\frac{2\pi}{a^{2}}Q_{X}L_{y}\frac{a}{r_{s}}}
                    |{\bf K}^{(m)},\sigma \rangle
  &=&|{\bf K}^{(m)},\sigma \rangle,
  \nonumber\\
  {\bf K}^{(m)}=(K_{x}^{(m)},K_{y}^{(m)})&=&
\left(\frac{2\pi m_{x}}{ L_{x}r_{s}a},\frac{2\pi r_{s}m_{y}}{a L_{y}}\right),
\end{eqnarray}
where $L_{x}, L_{y}$ are integers, $m_{x}=0,\cdots,L_{x}-1$ 
and $m_{y}=-[N_{e}L_{y}/2],\cdots,[N_{e}L_{y}/2]-1$ ( $[x]$ is the
integral part of $x$).
The thermodynamic limit $L_{x},L_{y},N_{e}\rightarrow \infty$ 
is taken in the following.
In this limit, the completeness  becomes
   \begin{eqnarray}
\label{eqn:completeness}
   \sum_{m,\sigma}|{\bf K}^{(m)},\sigma
       \rangle\langle {\bf K}^{(m)},\sigma|
 &=&\sum_{{\bf m},\sigma}
  |\frac{2\pi m_{x}}{L_{x}r_{s}a},\frac{2\pi r_{s}m_{y}}{ aL_{y}},\sigma
    \rangle\langle
 \frac{2\pi m_{x}}{ L_{x}r_{s}a},\frac{2\pi r_{s}m_{y}}{ aL_{y}},\sigma|
\nonumber\\
 &=&\sum_{\sigma}\frac{\cal A}{(2\pi)^{2}}
   \int_{0}^{2\pi/(r_{s}a)} dK_{x}
   \int_{-\infty}^{\infty}  dK_{y}
   |{\bf K},\sigma\rangle\langle{\bf K},\sigma|=1,
   \end{eqnarray}
   where ${\cal A} =L_{x}L_{y}a^{2}$.

\section{Current correlation function}
First, we represent a current correlation function 
in terms of the magnetic translational property of the striped Hall gas.
Next, the current correlation function is projected onto the $l$th LL,
and is evaluated by including the NG mode.
We  use the HFA and the SMA to evaluate the current correlation function.
\subsection{Symmetry of the  current correlation function}
We set $a=1$ in the following  calculation for the simplicity.
 The current correlation function  of the striped Hall gas 
 $|{\bf 0},\sigma_{0}\rangle$ is defined by
  \begin{eqnarray}
 \label{eqn:Fourier}
  \tilde{\Pi}^{\mu\nu}
  ({\bf k},{\bf k}^{\prime};\omega,\omega^{\prime})
  =\int dt dt^{\prime} d^{2}r d^{2}r^{\prime}
   \langle {\bf 0},\sigma_{0}|{\rm T}j^{\mu}({\bf r},t)
            j^{\nu}({\bf r}^{\prime},t^{\prime})|{\bf 0},\sigma_{0}\rangle
    e^{i{\bf k}\cdot {\bf r}
        +i{\bf k}^{\prime}\cdot {\bf r}^{\prime}
        -i\omega t -i\omega^{\prime}t^{\prime}},
  \end{eqnarray}
where ${\rm T}$ means the time-ordered product.
Let us write the current operator  as 
 $j^{\mu}({\bf r},t)=e^{iHt}
 j^{\mu}({\bf r},0)e^{-iHt}$.
By using the completeness of energy eigenstates 
Eq.~(\ref{eqn:completeness}) 
between current operators in 
Eq.~(\ref{eqn:Fourier}),
 the current correlation function becomes
\begin{eqnarray}
&&  \tilde{\Pi}^{\mu\nu}
  ({\bf k},{\bf k}^{\prime};\omega,\omega^{\prime})
=-i2\pi\delta(\omega+\omega^{\prime})
    K^{\mu\nu}_{\omega}({\bf k},{\bf k}^{\prime}),
\end{eqnarray}
where
\begin{eqnarray}
&& K^{\mu\nu}_{\omega}({\bf k},{\bf k}^{\prime})=
\nonumber\\
&&\sum_{\sigma}\frac{\cal A}{(2\pi)^{2}}
   \int_{0}^{2\pi/r_{s}} dK_{x}
   \int_{-\infty}^{\infty}  dK_{y}
\left\{
\frac{\langle 0,\sigma_{0}|\tilde{j}^{\mu}({\bf k},0)
  |{\bf K},\sigma\rangle\langle 
   {\bf K},\sigma|
\tilde{j}^{\nu}({\bf k}^{\prime},0)|0,\sigma_{0}\rangle}
     {\omega+E_{\sigma}({\bf K})-E_{\sigma_{0}}({\bf 0})-i\delta}
-\frac{\langle
0,\sigma_{0}|\tilde{j}^{\nu}({\bf k}^{\prime},0)|
           {\bf K},\sigma\rangle
   \langle {\bf K},\sigma|
   \tilde{j}^{\mu}({\bf k},0)|0,\sigma_{0}\rangle}
     {\omega-E_{\sigma}({\bf K})+E_{\sigma_{0}}({\bf 0})+i\delta}
\right\}\nonumber\\
\label{eqn:K}
\end{eqnarray}
and $\delta$ in the denominator is an infinitesimal positive constant.

Next, we transform $K^{\mu\nu}_{\omega}({\bf k},{\bf k}^{\prime})$
with the use of  Eq.~(\ref{eqn:ssb1}). 
The current operator is transformed under the magnetic translation
as 
$j^{\mu}({\bf r},0)=e^{-i2\pi r_{s}n_{x} Q_{Y}}
          j^{\mu}(\bar{x},y,0)
          e^{i2\pi r_{s}n_{x} Q_{Y}}$ and
$j^{\mu}(\bar{x},y,0)=e^{i2\pi y Q_{X}}j^{\mu}(\bar{x},0,0)
          e^{-i2\pi y Q_{X}}$,
where ${\bf r} =(r_{s}n_{x}+\bar{x}){\bf e}_{x}+y{\bf e}_{y}$ with 
$-r_{s}/2\leq\bar{x}\leq r_{s}/2$. 
${\bf e}_{x}$ and ${\bf e}_{y}$ are unit vectors in $x$ and $y$
direction, respectively.  
Inserting these current operators into Eq.~(\ref{eqn:K}),
we find that $Q_{X}$ and $Q_{Y}$ are replaced by eigenvalues of excited
 states  as follows,
\begin{eqnarray}
K^{\mu\nu}_{\omega}({\bf k},{\bf k}^{\prime})
&=&\sum_{\sigma}\frac{\cal A}{(2\pi)^{2}}
    \int_{0}^{2\pi/r_{s}} dK_{x}
\int_{-\infty}^{\infty} dK_{y}
\sum_{n_{x},n_{x}^{\prime}}
\int_{-r_{s}/2}^{r_{s}/2}d\bar{x}
d\bar{x}^{\prime}\int_{-\infty}^{\infty} dy  dy^{\prime}
e^{ik_{x}(r_{s}n_{x}+\bar{x})+ik_{y}y
  +ik_{x}^{\prime}(r_{s}n_{x}^{\prime}+\bar{x}^{\prime})
  +ik_{y}^{\prime}y^{\prime}}\nonumber\\
& &\times
\left\{
   \frac{\langle {\bf 0},\sigma_{0}|
         j^{\mu}(\bar{x},0,0)
         |{\bf K},\sigma\rangle
   \langle  {\bf K},\sigma|
          j^{\nu}(\bar{x}^{\prime},0,0)
          |{\bf 0},\sigma_{0}\rangle}
     {\omega+E_{\sigma}({\bf K})
            -E_{\sigma_{0}}({\bf 0})-i\delta}
    e^{ir_{s}K_{x}(n_{x}-n_{x}^{\prime})+iK_{y}(y-y^{\prime})}
\right.\nonumber\\
&&-\left.\frac{\langle {\bf 0},\sigma_{0}|
        j^{\nu}(\bar{x}^{\prime},0,0)
         |{\bf K},\sigma\rangle
         \langle  {\bf K},\sigma|
        j^{\mu}(\bar{x},0,0)
        |{\bf 0},\sigma_{0}\rangle}
     {\omega-E_{\sigma}({\bf K})
      +E_{\sigma_{0}}({\bf 0})+i\delta}
    e^{-ir_{s}K_{x}(n_{x}-n_{x}^{\prime})-iK_{y}(y-y^{\prime})}
\right\}\nonumber\\
&=&\sum_{\sigma}\frac{\cal A}{(2\pi)^{2}}
    \int_{0}^{2\pi/r_{s}} dK_{x}
\int_{-\infty}^{\infty}  dK_{y}
\int_{-r_{s}/2}^{r_{s}/2}d\bar{x}
d\bar{x}^{\prime}
 e^{ik_{x}\bar{x}+ik_{x}^{\prime}\bar{x}^{\prime}}
 \sum_{n_{x}^{\prime}} e^{i(k_{x}+k_{x}^{\prime})r_{s}n_{x}^{\prime}}
\int_{-\infty}^{\infty} dy^{\prime} e^{i(k_{y}+k_{y}^{\prime})y^{\prime}}
\nonumber\\
& &\times
\left\{
   \frac{\langle {\bf 0},\sigma_{0}|
         j^{\mu}(\bar{x},0,0)
         |{\bf K},\sigma\rangle
   \langle  {\bf K},\sigma|
          j^{\nu}(\bar{x}^{\prime},0,0)
          |{\bf 0},\sigma_{0}\rangle}
     {\omega+E_{\sigma}({\bf K})
            -E_{\sigma_{0}}({\bf 0})-i\delta}
\sum_{n_{x}^{\prime\prime}}
e^{i(k_{x}+K_{x})r_{s}n_{x}^{\prime\prime}}
\int_{-\infty}^{\infty} dy^{\prime\prime}
 e^{i(k_{y}+K_{y})y^{\prime\prime}}
    \right.\nonumber\\
& &\left.-\frac{\langle {\bf 0},\sigma_{0}|
        j^{\nu}(\bar{x}^{\prime},0,0)
         |{\bf K},\tau\rangle
         \langle  {\bf K},\sigma|
        j^{\mu}(\bar{x},0,0)
        |{\bf 0},\sigma_{0}\rangle}
     {\omega-E_{\sigma}({\bf K})
      +E_{\sigma_{0}}({\bf 0})+i\delta}
\sum_{n_{x}^{\prime\prime}}
e^{i(k_{x}-K_{x})r_{s}n_{x}^{\prime\prime}}
\int_{-\infty}^{\infty} dy^{\prime\prime} e^{i(k_{y}-K_{y})y^{\prime\prime}}
\right\},
\end{eqnarray}
where $y^{\prime\prime}=y-y^{\prime}$ 
and $n_{x}^{\prime\prime}=n_{x}-n_{x}^{\prime}$.
After carrying out 
  $y^{\prime}, y^{\prime\prime}$ integrations 
and  the summations
 of $n_{x}^{\prime}$ and $n_{x}^{\prime\prime}$,
the  delta functions  in the correlation functions appear as
\begin{eqnarray}
K^{\mu\nu}_{\omega}({\bf k},{\bf k}^{\prime})
&=&\sum_{\sigma}{\cal A}\frac{(2\pi)^{2}}{r_{s}^{2}}
\int_{-r_{s}/2}^{r_{s}/2}d\bar{x}
d\bar{x}^{\prime}
 \sum_{N,N^{\prime}}
\int_{0}^{2\pi/r_{s}} dK_{x}\int_{-\infty}^{\infty}  dK_{y}
 e^{ik_{x}\bar{x}+ik_{x}^{\prime}\bar{x}^{\prime}}
\delta({\bf k}+{\bf k}^{\prime}-\frac{2\pi {\bf N}}{r_{s}})
\nonumber\\
&&\times
\left\{   \frac{\langle {\bf 0},\sigma_{0}|
         j^{\mu}(\bar{x},0,0)
         |{\bf K},\sigma\rangle
   \langle  {\bf K},\sigma|
          j^{\nu}(\bar{x}^{\prime},0,0)
          |{\bf 0},\sigma_{0}\rangle}
     {\omega+E_{\sigma}({\bf K})
            -E_{\sigma_{0}}({\bf 0})-i\delta}
\delta ({\bf k}+{\bf K}-\frac{2\pi {\bf N}^{\prime}}{r_{s}})
   \right.\nonumber\\
&&- \left. 
     \frac{\langle {\bf 0},\sigma_{0}|
        j^{\nu}(\bar{x}^{\prime},0,0)
         |{\bf K},\sigma\rangle
         \langle  {\bf K},\sigma|
        j^{\mu}(\bar{x},0,0)
        |{\bf 0},\sigma_{0}\rangle}
     {\omega-E_{\sigma}({\bf K})
      +E_{\sigma_{0}}({\bf 0})+i\delta}
\delta({\bf k}-{\bf K}-\frac{2\pi {\bf N}^{\prime}}{r_{s}})
\right\}\nonumber\\
&=&\sum_{\sigma}{\cal A}\frac{(2\pi)^{2}}{r_{s}^{2}}
\int_{-r_{s}/2}^{r_{s}/2}d\bar{x}
d\bar{x}^{\prime}
 \sum_{N}\int d^{2}K
 e^{ik_{x}\bar{x}+ik_{x}^{\prime}\bar{x}^{\prime}}
\delta({\bf k}+{\bf k}^{\prime}-\frac{2\pi {\bf N}}{r_{s}})
\nonumber \\
&&\times
\left\{  \frac{\langle {\bf 0},\sigma_{0}|
         j^{\mu}(\bar{x},0,0)
         |{\bf K},\sigma\rangle
   \langle  {\bf K},\sigma|
          j^{\nu}(\bar{x}^{\prime},0,0)
          |{\bf 0},\sigma_{0}\rangle}
     {\omega+E_{\sigma}({\bf K})
            -E_{\sigma_{0}}({\bf 0})-i\delta}
 \delta({\bf K}+{\bf k})
\right.\nonumber\\
&&\left.  -   \frac{\langle {\bf 0},\sigma_{0}|
         j^{\nu}(\bar{x}^{\prime},0,0)
         |{\bf K},\sigma\rangle
         \langle  {\bf K},\sigma|
        j^{\mu}(\bar{x},0,0)
        |{\bf 0},\sigma_{0}\rangle}
      {\omega-E_{\sigma}({\bf K}) +E_{\sigma_{0}}({\bf 0})+i\delta}
\delta({\bf K}-{\bf k})
\right\},
\label{eqn:delta}
\end{eqnarray}
where ${\bf N}=N{\bf e}_{x}$, ${\bf N}^{\prime}=N^{\prime}{\bf e}_{x}$,
$N$ and $N^{\prime}$ are integers. 
In Eq~(\ref{eqn:delta}), we extend the finite $K_{x}$ integral
region into the infinite region with the use of an assumption
 $E_{\sigma}(K_{x}+2\pi/r_{s},K_{y})=E_{\sigma}({\bf K})$. 
After the  ${\bf K}$ integration,
the current correlation function becomes
\begin{eqnarray}
\label{eqn:CC}
 K^{\mu\nu}_{\omega}({\bf k},{\bf k}^{\prime})
=(2\pi)^{2}
   \sum_{N}\delta(\hat{\bf k}+\hat{\bf k}^{\prime}-2\pi {\bf N})
   \tilde{\Pi}^{\mu\nu}_{N}({\bf k},\omega),
\end{eqnarray}
where $\hat{{\bf k}}=(r_{s}k_{x},k_{y}/r_{s})$.
$\tilde{\Pi}^{\mu\nu}_{N}({\bf k},\omega)$
is defined as follows,
\begin{eqnarray}
\tilde{\Pi}^{\mu\nu}_{N}({\bf k},\omega)
&=&\frac{\cal A}{r_{s}^{2}}\sum_{\sigma} 
\int_{-r_{s}/2}^{r_{s}/2}d\bar{x}
\int_{-r_{s}/2}^{r_{s}/2}d\bar{x}^{\prime} 
 e^{ik_{x}\bar{x}+i(k_{x}-\frac{2\pi N}{r_{s}})\bar{x}^{\prime}}
\nonumber\\
&\times&\left\{
\frac{\langle {\bf 0},\sigma_{0}|
         j^{\mu}(\bar{x},0,0)
        |-{\bf k},\sigma\rangle   
         \langle -{\bf k},\sigma|
                   j^{\nu}(\bar{x}^{\prime},0,0)
          |{\bf 0},\sigma_{0}\rangle}
     {\omega+E_{\sigma}(-{\bf k})
            -E_{\sigma_{0}}({\bf 0})-i\delta}
  -\frac{\langle {\bf 0},\sigma_{0}|
        j^{\nu}(\bar{x}^{\prime},0,0)
         |{\bf k},\sigma\rangle
         \langle {\bf k},\sigma|
        j^{\mu}(\bar{x},0,0)
        |{\bf 0},\sigma_{0}\rangle}
     {\omega-E_{\sigma}({\bf k})
      +E_{\sigma_{0}}({\bf 0})+i\delta}
 \right\}.\nonumber\\
\end{eqnarray}
We use 
$j^{\mu}(\bar{x},0,0)=
\int d^{2}k^{\prime}\tilde{j}^{\mu}({\bf k}^{\prime},0)
e^{-ik_{x}^{\prime}\bar{x}}/(2\pi)^{2}$ 
and  $\int_{-r_{s}/2}^{r_{s}/2}d\bar{x}
e^{i(k_{x}-k_{x}^{\prime})\bar{x}}
=2\sin{((k_{x}-k_{x}^{\prime})r_{s}/2)}/(k_{x}-k_{x}^{\prime})$.
Then we obtain 
\begin{eqnarray}
\tilde{\Pi}^{\mu\nu}_{N}({\bf k},\omega)&=&
\label{eqn:correlation2}
4{\cal  A}\sum_{\sigma}
\int\frac{d^{2}k^{\prime}}{(2\pi)^{2}}
   \frac{\sin{((\hat{k}_{x}-\hat{k}_{x}^{\prime})/2})}
        {\hat{k}_{x}-\hat{k}_{x}^{\prime}}
\int\frac{d^{2}k^{\prime\prime}}{(2\pi)^{2}}
    \frac{\sin{((-\hat{k}_{x}-\hat{k}_{x}^{\prime\prime})/2+\pi N)}}
    {-\hat{k}_{x}-\hat{k}_{x}^{\prime\prime}+2\pi N}
\nonumber\\
&\times&\left\{
\frac{               \langle {\bf 0},\sigma_{0}|
            \tilde{j^{\mu}}({\bf k}^{\prime},0)
     |-{\bf k},\sigma\rangle     
     \langle  -{\bf k},\sigma|
            \tilde{j}^{\nu}({\bf k}^{\prime\prime},0) 
           |{\bf 0},\sigma_{0}\rangle}
    {\omega+E_{\sigma}(-{\bf k})
             -E_{\sigma_{0}}({\bf 0})-i\delta}
 -   \frac{           \langle {\bf 0},\sigma_{0}|
    \tilde{j^{\nu}}({\bf k}^{\prime\prime},0)
     |{\bf k},\sigma\rangle
     \langle {\bf k},\sigma|
                 \tilde{j}^{\mu}({\bf k}^{\prime},0)
            |{\bf 0},\sigma_{0}\rangle}
  {\omega -E_{\sigma}({\bf k})
            +E_{\sigma_{0}}({\bf 0})+i\delta}
\right\}.
\end{eqnarray}
The delta function of  ${\bf k}$ and
${\bf k}^{\prime}$ is periodic in $k_{x}$ direction
because of the periodicity of the striped Hall gas.
When ${\bf k}=0$, the cyclotron resonance is derived
with the use of Eq.~(\ref{eqn:correlation2}) in  Appendix B.
If the magnetic translational symmetries in
both  $x$ and $y$ direction were unbroken,
then the delta function would become $(2\pi)^{2}
\delta({\bf k}+{\bf k}^{\prime})$ and
there would be no  trigonometric factor.

\subsection{Evaluation of the current correlation function}
 We evaluate the current correlation function
 of the striped Hall gas in the half-filled $l$th LL.
 The striped Hall gas state spontaneously breaks the magnetic translational 
symmetry in $x$ direction and there exists  the NG mode. 
 We project the current correlation function onto the $l$th LL
to study the contribution of  the NG mode.
Since the species index $\sigma$ becomes $\sigma_{0}
$ after the LL projection,
we do not write the species index of the LL projected current
correlation function in the following in the present paper. 
 We use the UCDW in the HFA~\cite{HFref3,HFref4} for the
 ground state, and use the SMA ~\cite{SMA,SMA1,SMA2} for the NG 
 mode in order to evaluate the right hand side of 
Eq.~(\ref{eqn:correlation2}).

The UCDW in the HFA and the NG mode in the  SMA  are discussed 
by using the von Neumann lattice~(vNL) formalism,~\cite{vNL1}
in which the QHS is represented as a two-dimensional lattice system.
In the vNL formalism, the one-electron states are expanded by the vNL
basis. Let us introduce the vNL basis.
A discrete set of coherent states of guiding-center coordinates,
\begin{eqnarray}
(X+iY)|\alpha_{mn}\rangle=z_{mn}|\alpha_{mn}\rangle,
\end{eqnarray}
 is a complete 
set of the $(X,Y)$ space. Here we use $z_{mn}=(mr_{s}+in/r_{s})$
with integers $m$ and $n$.
These coherent states are localized at the position $(mr_{s},n/r_{s})$.
By Fourier transforming these states, we obtain the orthonormal basis in 
the momentum representation,
$|\beta_{{\bf p}}\rangle=\sum_{mn}e^{ip_{x}m+ip_{y}n}
                           |\alpha_{mn}\rangle/\beta({\bf p})$,
where $\beta({\bf p})=(2 {\rm Im} \tau)^{1/4}e^{i\tau p_{y}^{2}/4\pi}
  \vartheta_{1}((p_{x}+\tau p_{y})/2\pi|\tau)$.
The $\vartheta_{1}$ is a Jaccobi's theta function and $\tau=ir_{s}^{2}$.
The two-dimensional  momentum ${\bf p}$ is defined in the
  magnetic   Brillouin-zone (BZ), $|p_{i}|<\pi$.
A discrete set of  
the eigenstate of the one-particle free Hamiltonian,
\begin{eqnarray}
\frac{m\omega_{c}^{2}}{2}(\xi^{2}+\eta^{2})|f_{l}\rangle
=\omega_{c}(l+\frac{1}{2})|f_{l}\rangle,
l=0,1,2,\cdots,
\end{eqnarray}
is the complete set of the $(\xi,\eta)$ space.
The Hilbert space is spanned by the direct product
of these states, 
   $|l,{\bf p}\rangle=|f_{l}\rangle\otimes 
                               |\beta_{{\bf p}}\rangle$.
In the following in this paper, we do not write the time dependence
explicitly.
   A field operator of an electron is expanded by the vNL basis as
   \begin{eqnarray}
   \Psi ({\bf r})=\sum\limits_{l=0}^{\infty}\int_{\rm BZ}
   \frac{d^{2}p}{(2\pi)^{2}}
   b_{l}({\bf p})\langle {\bf r}|
                          l,{\bf p}\rangle,
   \end{eqnarray}
   where $b_{l}({\bf p})$ is the  anti-commuting annihilation operator
which obeys
    \begin{eqnarray}
   \label{eqn:anti}
   \{b_{l}({\bf p}),b_{l^{\prime}}^{\dag}({\bf p}^{\prime})\}
    =\delta_{l l^{\prime}}\sum_{N}(2\pi)^{2}
     \delta({\bf p}-{\bf p}^{\prime}-2\pi {\bf n})
e^{i\phi (p^{\prime},n)}
   \end{eqnarray} 
with  a boundary condition   
   $b_{l}({\bf p}+2\pi {\bf n})=
   e^{i\phi (p,n)}b_{l}({\bf p})$. 
Here,   ${\bf n}=(n_{x},n_{y})$, $n_{x}$ and $n_{y}$ are integers and 
   $\phi (p,n)=\pi(n_{x}+n_{y})-n_{y}p_{x}$.

  The mean-field state of the striped Hall gas in the  HFA
 has an anisotropic Fermi sea (FIG.1) 
  which is uniform in $p_{x}$ direction.
The mean-field state is constructed as 
      $|0\rangle=N_{1}\Pi_{{\bf p}\in {\rm FS}}b^{\dag}_{l}({\bf p})
         |vac\rangle$.
     Here $N_{1}$ is the normalization constant,  
     ${\rm FS}$ denotes the Fermi sea and
   $|vac\rangle$ is the vacuum state which is occupied up to 
   the $l-1$th LL.
\begin{figure}
\includegraphics[width=4.5cm,clip]{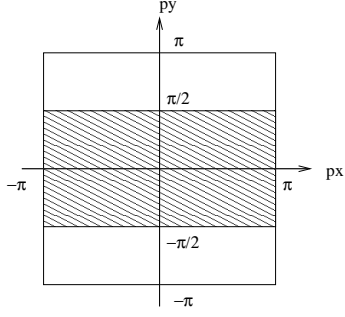}
\caption{Fermi sea in the magnetic  BZ of $l$th Landau level}
\end{figure}
   The two point function is given by
   \begin{eqnarray}
    \label{eqn:twopoint}
    \langle 0|b^{\dag}_{l}({\bf p})
            b_{l^{\prime}}({\bf p}^{\prime})|0\rangle
   =\delta_{ll^{\prime}}\theta\bigl[\mu-\epsilon_{\rm HF}^{(l)}
   ({\bf p})\bigr]
     \sum_{N}(2\pi)^{2}\delta
    ({\bf p}-{\bf p}^{\prime}
                 +2\pi{\bf N})e^{i\phi(p,N)},
   \end{eqnarray}
   where $\mu$ is a chemical potential and
   $\epsilon_{\rm HF}^{(l)}({\bf p})$ is 
a one-electron energy in the $l$th LL.
   The HF  state $|0\rangle$  is an eigenstate of 
   magnetic translation operators 
   with an eigenvalue ${\bf K}={\bf 0}$.

The SMA is a variational method to calculate the lowest excited state, 
and is  consistent with the GRPA in numerical
calculations.~\cite{phonon7}  
The low energy excitation in the SMA is a fluctuation of 
the density in the striped Hall gas.
Given the ground state $|0\rangle$, the NG mode in the SMA  at ${\bf k}$ 
is assumed as
\begin{eqnarray}
|{\bf k}\rangle 
&=&\frac{\tilde{\rho}_{\ast}({\bf k})}
{\sqrt{N_{e}^{\ast}s({\bf k})}}|0\rangle
\end{eqnarray}
with  the projected density operator,
\begin{eqnarray}
\label{eqn:rho}
\tilde{\rho}_{\ast}({\bf k})
=P_{l}\int d^{2}r \Psi^{\dag}({\bf r})e^{i{\bf k}\cdot{\bf X}}
\Psi({\bf r})P_{l}
=\int_{\rm BZ}\frac{d^{2}p}{(2\pi)^{2}}b^{\dag}_{l}({\bf p})
   b_{l}({\bf p}-\hat{{\bf k}})
  e^{-i\frac{\hat{k}_{x}}{4\pi}(2p_{y}-\hat{k}_{y})},
\end{eqnarray}
where $P_{l}$ is a   projection operator onto the subspace of the $l$th LL.
$N_{e}^{\ast}$ is the electron number in the $l$th LL, and the function
$s({\bf k})$, defined as
$s({\bf k})=\langle 0|
                       \tilde{\rho}_{\ast}(-{\bf k})
                       \tilde{\rho}_{\ast}({\bf k})|0\rangle/N_{e}^{\ast}$,
is the static structure function.
The excitation energy in the SMA is given by
$E_{\rm SMA}({\bf k})=f({\bf k})/s({\bf k})$,
where  $f({\bf k})=\langle 0|
[\tilde{\rho}_{\ast}(-{\bf k}),[H^{(l)},\tilde{\rho}_{\ast}({\bf k})]]
|0\rangle /2N_{e}^{\ast}$.
The operator $H^{(l)}=\int d^{2}k\tilde{\rho}_{\ast}({\bf k})
v_{l}(k)\tilde{\rho}_{\ast}({\bf k})/(2(2\pi)^{2})$ 
is the Hamiltonian projected onto the $l$th LL,
where 
$v_{l}({\bf k})=\exp{(-{\bf k}^{2}/{4\pi})}
[L_{l}({\bf k}^{2}/{4\pi})]^{2} 2\pi q^{2}/{k}$,
and $L_{l}(x)$ is the Laguerre polynomial.   
From our previous paper,~\cite{SMA}  we know 
$E_{\rm SMA}({\bf k})
=q^{2}|\hat{k}_{y}|\{A_{0}\hat{k}_{x}^{2}+B_{0}\hat{k}_{y}^{4}
+O(k_{x}^{2}k_{y}^{2},k_{x}^{4})\}$
and that the NG mode in the SMA  is an 
eigenstate of magnetic translation operators with an eigenvalue 
${\bf K}={\bf k}$.

In terms of $|0\rangle$ and $|{\bf k}\rangle$,
the matrix element  in the integrand of Eq.~(\ref{eqn:correlation2}),
i.e., $\langle {\bf 0},\sigma_{0}|
            \tilde{j^{\mu}}({\bf k}^{\prime})
     |{\bf k},\sigma\rangle$, 
is replaced by  
\begin{eqnarray}
\langle 0| P_{l} \tilde{j^{\mu}}({\bf k}^{\prime})P_{l}
     |{\bf k}\rangle
=\langle
0|\alpha^{\mu}({\bf k}^{\prime})
  \tilde{\rho}_{\ast}({\bf k}^{\prime})|{\bf k}\rangle
=\frac{\alpha^{\mu}({\bf k}^{\prime})}
{\sqrt{N_{e}^{\ast}s({\bf k})}}
\langle 0|\tilde{\rho}_{\ast}({\bf k}^{\prime})
          \tilde{\rho}_{\ast}({\bf k})|0\rangle.
\end{eqnarray}
Here, $\alpha^{\mu}({\bf k}^{\prime})$ is defined 
by 
  \begin{equation}
\alpha^{\mu}({\bf k}^{\prime})=
      \begin{cases}
      e^{\frac{-{\bf k}^{\prime 2}}{8\pi}}
            L_{l}(\frac{{\bf k}^{\prime 2}}{4\pi})&
       \text{; $\mu=0$}\\     
      -\sum_{j=x,y}
       i\omega_{c}\epsilon^{ij}\frac{\partial}{\partial k^{\prime}_{j}}
            \bigl\{
            e^{\frac{-{\bf k}^{\prime2}}{8\pi}}
            L_{l}(\frac{{\bf k}^{\prime}}{4\pi})\bigr\}
      =-\sum_{j=x,y}
        i\omega_{c}e^{\frac{-{\bf k}^{\prime 2}}{8\pi}}
        \epsilon^{ij}\frac{k^{\prime}_{j}}{2\pi} 
       \Bigl\{L_{l}^{\prime}(\frac{{\bf k}^{\prime}}{4\pi})
            -\frac{1}{2}L_{l}(\frac{{\bf k}^{\prime}}{4\pi})\Bigr\}&
           \text{; $\mu=i$}.
      \end{cases}
  \end{equation}
From  Eqs.~(\ref{eqn:anti}), (\ref{eqn:twopoint}) and
(\ref{eqn:rho}), the expectation value of 
$\tilde{\rho}_{\ast}({\bf k}^{\prime})\tilde{\rho}_{\ast}({\bf k})$
 is obtained as follows, 
\begin{eqnarray}
  \label{eqn:product}
\langle 0|\tilde{\rho}_{\ast}({\bf k}^{\prime})
           \tilde{\rho}_{\ast}({\bf k})|0\rangle
&=&\int_{\rm BZ}\frac{d^{2}p}{(2\pi)^{2}}
   \int_{\rm BZ}\frac{d^{2}p^{\prime}}{(2\pi)^{2}}
   e^{-i\frac{\hat{k}^{\prime}_{x}}{4\pi}
            (2p_{y}-\hat{k}^{\prime}_{y})
      -i\frac{\hat{k}_{x}}{4\pi}
            (2p_{y}^{\prime}-\hat{k}_{y}) }
   \langle 0|b^{\dag}_{l}({\bf p})b_{l}({\bf p}-\hat{{\bf k}}^{\prime})
    b_{l}^{\dag}({\bf p}^{\prime})b_{l}({\bf p}^{\prime}-\hat{{\bf k}})
    |0\rangle
\nonumber\\
&=&\int_{\rm BZ}\frac{d^{2}p}{(2\pi)^{2}}
   \int_{\rm BZ}\frac{d^{2}p^{\prime}}{(2\pi)^{2}}
   e^{-i\frac{\hat{k}^{\prime}_{x}}{4\pi}
            (2p_{y}-\hat{k}^{\prime}_{y})
      -i\frac{\hat{k}_{x}}{4\pi}
            (2p_{y}^{\prime}-\hat{k}_{y}) }
\{-\langle 0|b^{\dag}_{l}({\bf p})b_{l}^{\dag}({\bf p}^{\prime})
 b_{l}({\bf p}-\hat{{\bf k}}^{\prime})
 b_{l}({\bf p}^{\prime}-\hat{{\bf k}})|0\rangle
\nonumber\\
&&+\sum_{{\bf n}}(2\pi)^{2}\delta({\bf p}+{\bf k}^{\prime}
             -{\bf p}^{\prime}-2\pi{\bf n})e^{i\phi(p^{\prime},n)}
\langle 0|b^{\dag}_{l}({\bf p})b_{l}({\bf p}^{\prime}-\hat{{\bf k}})
|0\rangle
\}\nonumber\\
&=&\langle\tilde{\rho}_{\ast}({\bf k}^{\prime})\rangle
   \langle\tilde{\rho}_{\ast}({\bf k})\rangle
  + e^{i\frac{\hat{k}_{x}^{\prime} \hat{k}_{y}^{\prime}}{4\pi}
              -i\frac{\hat{k}_{x}\hat{k}_{y}}{4\pi}}
     \sum_{\bf n}(2\pi)^{2}\delta(\hat{{\bf k}}
                      +\hat{{\bf k}}^{\prime}-2\pi{\bf n})
\gamma (n_{x},k_{y})e^{i\pi n_{x}}\delta_{n_{y}0}.
  \end{eqnarray}
Here $\langle 0|b^{\dag}_{l}({\bf p})b_{l}^{\dag}({\bf p}^{\prime})
 b_{l}({\bf p}-\hat{{\bf k}}^{\prime})
 b_{l}({\bf p}^{\prime}-\hat{{\bf k}})|0\rangle$ is decomposed into
the direct term and  the exchange term, and the product of
$ \langle\tilde{\rho}_{\ast}({\bf k}^{\prime})\rangle $
comes from the direct term.
The $\gamma (n_{x},k_{y})$ is defined as 
\begin{eqnarray}
\gamma (n_{x},k_{y})
=     \delta_{n_{x}0}\frac{|\hat{k}_{y}|}{2\pi} 
   +\frac{1}{n_{x}\pi}\left\{\sin\left(\frac{n_{x}\pi}{2}\right)
   -e^{i\frac{n_{x}\hat{k}_{y}}{2}}
       \sin\left(\frac{n_{x}}{2}(\pi-|\hat{k}_{y}|)\right)
  \right\}(1-\delta_{n_{x}0}).
\end{eqnarray}
 The expectation value 
$ \langle \tilde{\rho}_{\ast}
   ({\bf k})\rangle  =
    \langle 0|\tilde{\rho}_{\ast}
    ({\bf k})   |0\rangle  
  =  (2\pi)^{2}
     \sum_{n_{x}}\delta(\hat{k}_{x}+2\pi n_{x})
                       \delta(\hat{k}_{y})\exp({i\pi n_{x}})
  \sin{(n_{x}\pi/2)}/(n_{x}\pi)$
vanishes in the small ${\bf k}$ region (${\bf k}\neq 0$).
Thus the matrix element in the numerator in  Eq.~(\ref{eqn:correlation2}) 
in the small ${\bf k}$ region becomes
\begin{eqnarray}
\langle 0|
            P_{l}\tilde{j^{\mu}}({\bf k}^{\prime})P_{l}
           |{\bf k}\rangle
          \langle {\bf k}|
            P_{l}\tilde{j^{\nu}}({\bf k}^{\prime\prime})P_{l}|0\rangle
&=&\frac{\alpha^{\mu}({\bf k}^{\prime})
       \alpha^{\nu}({\bf k}^{\prime\prime})}
      {N_{e}^{\ast}s({\bf k})}
e^{i\frac{(\hat{k}_{x}^{\prime} \hat{k}_{y}^{\prime}
  -\hat{k}_{x}^{\prime\prime}   \hat{k}_{y}^{\prime\prime})}{4\pi}}
 \sum_{n_{x},{\bf n}^{\prime}}
\gamma (n_{x},k_{y})\gamma (n_{x}^{\prime},k_{y})
e^{i\pi  (n_{x}+n_{x}^{\prime})}\delta_{n_{y}0}\delta_{n_{y}^{\prime}0}
\nonumber\\
&&\times
      (2\pi)^{2}\delta(\hat{\bf k}+\hat{\bf k}^{\prime}-2\pi{\bf n})
        (2\pi)^{2}\delta(-\hat{\bf k}+\hat{\bf k}^{\prime\prime}-2\pi
                    {\bf n}^{\prime}).   
\label{eqn:matrix}
\end{eqnarray}

By inserting Eq.~(\ref{eqn:matrix}) into 
 Eq.~(\ref{eqn:correlation2}) and
carrying out the 
${\bf k}^{\prime}$, ${\bf k}^{\prime\prime}$ integrations,
 the current correlation function
 $\tilde{\Pi}_{N\ast}^{\mu\nu}({\bf k},\omega)$  in the $l$th LL is
obtained as
\begin{eqnarray}
&&
\sum_{n_{x},n_{x}^{\prime}}
 \frac{(-1)^{N}{\cal A}
  \alpha^{\mu}(-k_{x}+2\pi n_{x}/r_{s},-k_{y})
   \alpha^{\nu}(k_{x}+2\pi n_{x}^{\prime}/r_{s},k_{y})
   \gamma (n_{x},k_{y})\gamma (n_{x}^{\prime},k_{y})
   \sin ^{2}\hat{k}_{x}}
      {N_{e}^{\ast}s({\bf k})
      (\hat{k}_{x}-\pi n_{x})
      (\hat{k}_{x}+\pi n_{x}^{\prime}-\pi N)}
\nonumber\\
&&\times
\left\{
\frac{e^{-i(n_{x}+n_{x}^{\prime})\hat{k}_{y}/2}}
    {\omega +E_{\rm SMA}(-{\bf k})-i\delta}
-\frac{e^{i(n_{x}+n_{x}^{\prime})\hat{k}_{y}/2}}
        {\omega -E_{\rm SMA}({\bf k})+i\delta}
\right\}.
\end{eqnarray}
The dominant contribution in the current correlation function
$\tilde{\Pi}_{0\ast}^{\mu\nu}({\bf k},\omega)$ comes from
the  $n_{x}=n_{x}^{\prime}=0$  term
at the small ${\bf k}$ region 
due to the factors  $(a\hat{k}_{x}-\pi n_{x})^{-1}$ and 
$(a\hat{k}_{x}+\pi n_{x}^{\prime}-\pi N)^{-1}$.
The dominant term of 
 $\tilde{\Pi}_{0\ast}^{\mu\nu}({\bf k},\omega)$ reads
\begin{eqnarray}
\label{eqn:result2}
 &&{\cal A}
          \left(\frac{|\hat{k}_{y}|}{2\pi}
     \frac{\sin{\hat{k}_{x}}}{\hat{k}_{x}}\right)^{2}
     \frac{\alpha^{\mu}(-{\bf k})\alpha^{\nu}({\bf k})}
          {N_{e}^{\ast}s({\bf k})}
     \left\{\frac{1}{\omega +E_{\rm SMA}(-{\bf k})-i\delta}
          - \frac{1}{\omega -E_{\rm SMA}({\bf k}) +i\delta}
     \right\}\nonumber\\
  &=&\alpha^{\mu}(-{\bf k})\alpha^{\nu}({\bf k})
      \frac{|\hat{k}_{y}|}{2\pi}
      \left\{\frac{1}{\omega +E_{\rm SMA}(-{\bf k})-i\delta}
          - \frac{1}{\omega -E_{\rm SMA}({\bf k})+i\delta}
     \right\}.
\end{eqnarray}
Here we use the static structure factor 
 $s({\bf k})
    =|\hat{k}_{y}|/(2\pi\nu_{\ast})$ at a small ${\bf k}$, 
where  $\nu_{\ast}= N_{e}^{\ast}/{\cal A}$  is a filling
 factor of the $l$th LL.
We also use   $\lim_{x\to 0}\sin{x}/x=1$. 

\section{Implications of the NG mode}
From the current correlation function of the striped Hall gas,  we 
see  implications of the NG mode.
In the subsection A, we find  a sharp energy absorption  
at the NG mode frequency 
besides at the cyclotron frequency.
In the subsection B, we study
the dominant resonance of the density correlation function  
in the long wavelength limit 
and show that the density correlation function
satisfies the f-sum rule.

\subsection{Photon energy absorption rate} 

When an  external electromagnetic wave is added,
the striped Hall gas absorbs the photon energy.
We study the photon  energy absorption rate 
which is proportional to the current correlation
function, and inversely proportional to the frequency
 $\omega$~(see appendix A).
A sharp absorption  occurs  at $\omega=\omega_{c}$
 when ${\bf k}=0$~(see appendix B). 
The striped Hall gas has the  NG mode $\omega=\omega_{NG}({\bf k})$,
and a sharp absorption  may also occur at the NG mode frequency 
which vanishes in the $|{\bf k}|\rightarrow 0$ limit.

 As a typical example, we assume that the electromagnetic wave is propagating 
in the three-dimensional
space~$({\bf r},z)$ as ${\bf E}={\cal E}_{L}({\bf e}_{x}+i{\bf
 e}_{y})e^{i{\bf k}\cdot{\bf r}+ik_{z}z-i\omega t}$, where 
${\cal  E}_{L}({\bf e}_{x}+i{\bf e}_{y})$ is a left-handed
circularly polarization vector, and $k_{z}$ is a wave number vector in
$z$ direction. 
The photon energy absorption rate of the striped Hall gas
 is expressed by the current correlation function as 
\begin{eqnarray}
P=\frac{2e^{2}{\cal E}_{L}^{2}}{\omega }
\bigl(
{\rm Im}_{\omega >0}\tilde{\Pi}_{0}^{xx}({\bf k},\omega)+
{\rm Im}_{\omega >0}\tilde{\Pi}_{0}^{yy}({\bf k},\omega)
\bigr),
\end{eqnarray}
which is derived in the appendix A.

In order to evaluate the contribution of the NG mode,
 we project the photon energy absorption rate onto
the $l$th LL as 
\begin{eqnarray}
P_{\rm NG}&=&
\frac{2e^{2}{\cal E}_{L}^{2}}{\omega }
\bigl(
{\rm Im}_{\omega >0}\tilde{\Pi}_{0\ast}^{xx}({\bf k},\omega)+
{\rm Im}_{\omega >0}\tilde{\Pi}_{0\ast}^{yy}({\bf k},\omega)
\bigr).
\end{eqnarray}
By inserting Eq.~(\ref{eqn:result2}) into the above equation, 
we compute the photon energy absorption rate as 
\begin{eqnarray}
P_{\rm NG}&=&\frac{e^{2}{\cal E}_{L}^{2}}{\omega }
    \sum_{i=x,y}\alpha^{i}(-{\bf k})\alpha^{i}({\bf k})
      \frac{|\hat{k}_{y}|}{2\pi}
      2\pi\delta(\omega -E_{\rm SMA}({\bf k}))
       \nonumber\\
&=&\left(\frac{e{\cal E}_{L}\omega_{c}(l+\frac{1}{2})}{2\pi}\right)^{2}
\frac{k_{x}^{2}+k_{y}^{2}}
{q^{2}(A_{0}\hat{k}_{x}^{2}+B_{0}\hat{k}_{y}^{4})}
 \delta(\omega 
-q^{2}|\hat{k}_{y}|(A_{0}\hat{k}_{x}^{2}+B_{0}\hat{k}_{y}^{4})),
\end{eqnarray}
where we use $E_{SMA}({\bf k})
=q^{2}|\hat{k}_{y}|(A_{0}\hat{k}_{x}^{2}+B_{0}\hat{k}_{y}^{4})$,
and
$\alpha^{i}({\bf k})=i\omega_{c}(l+1/2)\epsilon^{ij}k_{j}/(2\pi)$
in the small ${\bf k}$ region.
We also used  $e^{-x}=1$, $L_{l}(x)=1$,  
$L_{l}^{\prime}(x)=-l$ in $x\rightarrow 0$ limit.   
The ratio $(k_{x}^{2}+k_{y}^{2})/
(A_{0}\hat{k}_{x}^{2}+B_{0}\hat{k}_{y}^{4})$  converges 
to a constant value
when $k_{x}$ and $k_{y}$ approach zero in the same order 
or when $k_{y}$ approaches zero in the higher order than $k_{x}$.
In particular when 
we fix $k_{x}$ and take a $k_{y}\rightarrow 0$,
the photon energy absorption rate is obtained as follows, 
\begin{eqnarray}
P_{\rm NG}&=&\left(\frac{e{\cal
	  E}_{L}\omega_{c}(l+\frac{1}{2})}{2\pi}\right)^{2}
          \left(q^2 A_{0}r_{s}^{2}\right)^{-1}\delta(\omega).
\end{eqnarray}
This result shows  a sharp energy absorption at the NG mode frequency.
The same result is obtained for the right-handed circularly polarized
electromagnetic wave.
On the other hand, a sharp energy absorption at $\omega=\omega_{c}$
 with ${\bf k}={\bf 0}$ is derived by
using the current correlation function without the LL projection.
The result becomes the following~(see appendix B); 
\begin{eqnarray}
P_{\rm cyclotron}=\frac{(e{\cal
 E}_{L})^{2}\rho_{e}}{m}2\pi\delta(\omega-\omega_{c}),
\end{eqnarray} 
where $\rho_{e}={\rm N}_{e}/{\cal A}$.
The cyclotron resonance is not affected by the NG mode. 
A sharp absorption at
$\omega=E_{\rm SMA}({\bf k})$ occurs at a small ${\bf k}$
in addition to the sharp absorption at the cyclotron frequency.
For the right-handed circularly polarized electromagnetic wave,
$P_{\rm cyclotron}=0$.

The strength of  resonant peaks depends on a magnetic field $B$.
Fig.~2 shows the $B$-dependence of the photon energy absorption rate.
$P_{\rm NG}$ and $P_{\rm cyclotron}$ of the striped Hall gas depend
 on $B$ through the cyclotron frequency  $\omega_{c}$ 
and $a=\sqrt{2\pi/eB}$ which has been set $1$ so far.
The photon energy absorption rate at  the NG mode frequency 
in the SI units is 
$\lim_{E_{\rm SMA}\rightarrow 0} P_{\rm NG}/({\cal E}_{L}^{2}
\delta(\omega -E_{\rm SMA}))
=1.810\times 10^{9} B^{3/2}[{\rm s}^{2}~{\rm A}^{2}/{\rm kg}~{\rm m}]$
with $l=2$, $r_{s}=2.474$ and $A_{0}=0.351$. 
The  value of $r_{s}$ is determined by minimizing the HF energy per
 unit area.
The photon energy absorption rate at the cyclotron frequency in the SI units is
$P_{\rm cyclotron}/({\cal E}_{L}^{2}\delta(\omega-\omega_{c}))
=1.572 \times 10^{9}B
[{\rm s}^{2}~{\rm A}^{2}/{\rm kg}~{\rm m} ]$ 
with $l=2$.
The photon energy absorption rate
 at the  NG mode frequency is larger than the photon energy absorption rate
 at the cyclotron frequency.
\begin{figure}
\includegraphics[width=7cm,clip]{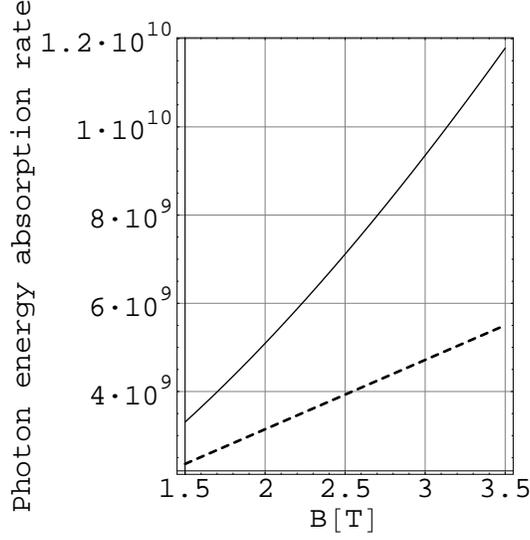}
\caption {$B$-dependence of the photon energy absorption rate 
with  $r_{s}=2.474$
at $l=2$. The photon absorption rate 
$P_{\rm NG}/({\cal E}_{L}^{2}
\delta(\omega - E_{\rm SMA}))$ at the small ${\bf k}$ is
 shown by a solid curve,
 and $P_{\rm cyclotron}/({\cal E}_{L}^{2}\delta(\omega-\omega_{c}))$ 
is shown by a dashed
 curve. We fix a filling factor as $\nu=2.5$ and change a magnetic field $B[T]$
 around $2.5[T]$ where an anisotropic resistivity at $\nu
 =2.5$ is observed. The unit of the vertical axis is 
$[{\rm s}^{2}~{\rm A}^{2}/{\rm kg}~{\rm m}^{2}]$.}
\end{figure}

The strength of $P_{\rm NG}$ depends on
 the incident angle of the electromagnetic wave. 
When the electromagnetic wave is tilted parallel to the stripes,
the photon energy absorption rate diverges in the $|{\bf k}|\rightarrow
0$ limit.
When the electromagnetic wave is tilted perpendicular to the
stripes, the photon energy absorption rate becomes finite. 
The anisotropy of $P_{NG}$ is a new property of the striped Hall gas.

\subsection{Properties of the density correlation function}
In this subsection, we discuss  properties of the density
correlation function which is a temporal diagonal element
 of the  current correlation function in the QHS.

When excitations have finite energy gaps,
the dominant resonance of the density correlation function
$K^{00}_{\omega}({\bf k},-{\bf k})$ 
occurs at $\omega=\omega_{c}$ in the long wavelength limit.
Since the striped Hall gas has a zero-energy excitation,
 there is also a
resonance of the density correlation function at the  NG mode frequency.
We study the NG mode resonance of the density correlation function in
the long wavelength limit and find that the dominant resonance occurs
at the NG mode frequency.

First  
we derive the cyclotron resonance in the density correlation function
when there is a finite energy gap.
The ground state and an excited state are denoted by $|0\rangle$ and
$|m\rangle$, respectively.
The current conservation law gives 
$\langle m|[H,\tilde{j^{0}}({\bf k})]|0\rangle
=i\sum_{i=x,y}k_{i}\langle m|\tilde{j}^{i}({\bf k})|0\rangle$,
where $\tilde{j^{\mu}}({\bf k})
=\int d^{2}r j^{\mu}({\bf r})
e^{i{\bf k}\cdot {\bf r}}$. 
Thus a matrix element of the density operator 
 $\langle m|\tilde{j^{0}}({\bf k})|0\rangle$ is
represented by
  $i\sum_{i=x,y}k_{i}\langle 0|\tilde{j}^{i}({\bf k})|m\rangle/(E_{m}-E_{0})$.
By using this matrix element and $\tilde{j}^{i}({\bf k})=\pi^{i}/m
+O({\bf k})$,
we can  rewrite the density correlation
function as follows,
  \begin{eqnarray}
 K_{\omega}^{00}({\bf k},-{\bf k})
&=& \sum_{m}\left\{
    \frac{\langle 0|\tilde{j}^{0}({\bf k})|m\rangle
     \langle  m|\tilde{j}^{0}(-{\bf k})|0\rangle}
         {\omega+E_{m}-E_{0}-i\delta}
   -\frac{\langle 0|\tilde{j}^{0}(-{\bf k})|m\rangle
     \langle  m|\tilde{j}^{0}({\bf k})|0\rangle}
         {\omega -E_{m}+E_{0}+i\delta}    
   \right\}
\nonumber\\
&=&\frac{1}{m^{2}}\sum_{i,j=x,y}\sum_{m}\frac{k_{i}k_{j}}{(E_{m}-E_{0})^{2}}
\left\{
 \frac{\langle 0|\pi^{i}|m\rangle
     \langle  m|\pi^{j}|0\rangle}
         {\omega+E_{m}-E_{0}-i\delta}
   -\frac{\langle 0|\pi^{j}|m\rangle
     \langle  m|\pi^{i}|0\rangle}
         {\omega -E_{m}+E_{0}+i\delta}+O({\bf k})    
\right\}
\nonumber\\
&=&\frac{N_{e} {\bf k}^{2}}{2\pi}
\left\{
\frac{1}{\omega+\omega_{c}-i\delta}-\frac{1}{\omega-\omega_{c}+i\delta}
\right\}+\frac{1}{m^{2}}\sum_{m}\frac{1}{(E_{m}-E_{0})^{2}}
O({\bf k}^{3}), 
\label{eqn:Kohnden}
\end{eqnarray}
where the operator $\pi^{i}$ is the covariant momentum defined as
\begin{eqnarray}
\pi^{x}&=&-\int d^{2}r\Psi^{\dag}({\bf r}) 2\pi\eta
          \Psi({\bf r}),
\nonumber\\
\pi^{y}&=&\int d^{2}r\Psi^{\dag}({\bf r})2\pi\xi
          \Psi({\bf r}).
\end{eqnarray}
When there is a finite energy gap, the
${\bf k}$-dependence of the energy difference
in the second term is negligible. Thus the  largest residue of the 
density correlation function
$O({\bf k}^{2})$ appears at the cyclotron frequency, 
and the  residues of the  density correlation function  at 
 the other excitations are $O({\bf k}^{3})$.   
The  cyclotron resonance is dominant at a small
${\bf k}$.~\cite{Zhang}

Next we discuss a resonance in the density correlation function
when there is a gapless excitation.
In this case, we cannot neglect the
${\bf k}$-dependence of the energy difference. 
Since the energy difference 
$E_{m}-E_{0}$ approaches zero as ${\bf k}$ approaches zero
and   cancels  $k_{i}$ in the numerator of 
Eq.~(\ref{eqn:Kohnden}), 
 residues of the density correlation function at gapless excitations 
in the second term of Eq.~(\ref{eqn:Kohnden}) may become larger than
$O({\bf k}^{2})$. 
We use the LL projected density correlation function
of the striped Hall
gas obtained in the previous section,
and   study a residue of the density correlation function at  
the NG mode  frequency.
Inserting  Eq.~(\ref{eqn:result2}) into  Eq.~(\ref{eqn:CC}),
we obtain  the long wavelength
limit of the density correlation function in the $l$th LL  as
\begin{eqnarray}
K_{\omega\ast}^{00}({\bf k},-{\bf k})&=&
(2\pi)^{2}\delta(0)^{2}\tilde{\Pi}^{00}_{0\ast}({\bf k},\omega)
\nonumber\\
&=&
{\cal A}\frac{|\hat{k}_{y}|}{2\pi}
\left\{
\frac{1}{\omega+E_{\rm SMA}({\bf k})-i\delta}
-\frac{1}{\omega -E_{\rm SMA}({\bf k})+i\delta}
\right\},
\label{eqn:KD2}
\end{eqnarray}
where we use $\alpha^{0}({\bf k})=1$ in the small ${\bf k}$ region.
By comparing Eq.~(\ref{eqn:Kohnden}) with  Eq.~(\ref{eqn:KD2}), 
we know that 
 the  residue of the  density correlation function  at the NG mode  frequency
 is ${\cal A} |\hat{k}_{y}|/(2\pi )$, and the residue of the density 
correlation
function at the cyclotron frequency is $N_{e} {\bf k}^{2}/2\pi$.  
The density correlation function of the
striped Hall gas has the largest residue at the  NG mode frequency 
in the long wavelength limit.
Hence, the dominant resonance of the density correlation function
occurs at the NG mode frequency.

The $\tilde{\Pi}^{00}_{0*}({\bf k},\omega)$  should satisfy
the f-sum rule~\cite{mahan} in the subspace of the $l$th LL
as the following,
\begin{eqnarray}
\int_{0}^{\infty}d\omega\omega {\rm Im}
\tilde{\Pi}_{0\ast}^{00}({\bf k},\omega)
=\pi\nu_{\ast}f({\bf k}).
\label{eqn:f-sum}
\end{eqnarray}
The left hand side is calculated in the SMA by using 
Eq.~(\ref{eqn:result2}), and is given by
\begin{eqnarray}
\int_{0}^{\infty}d\omega E_{\rm SMA}({\bf k})
\frac{|\hat{k}_{y}|}{2\pi}\pi\delta(\omega -E_{\rm SMA}({\bf k})).
\end{eqnarray}
Using  $E_{\rm SMA}({\bf k})=f({\bf k})/s({\bf k})$ and 
$s({\bf k})=|\hat{k}_{y}|/(2\pi\nu_{\ast})$, this becomes
$\pi \nu_{\ast}f({\bf k})$ which is the right hand side of
Eq.~(\ref{eqn:f-sum}). Hence, $\tilde{\Pi}_{0\ast}^{00}({\bf k},\omega)$
in the SMA satisfies the f-sum rule. This suggests that   
the SMA is reasonable.

\section{summary and discussion }
In this paper, we have studied the contribution of the NG mode 
at $\omega=\omega_{NG}({\bf k})$ to 
the current correlation function of the striped Hall gas.
 The striped Hall gas is the  UCDW state 
which is one of the HFA solutions and has an anisotropic Fermi surface. 
The striped Hall gas spontaneously breaks 
the magnetic translational symmetry and the magnetic rotational symmetry.
The symmetry breaking causes a gapless NG mode.

We evaluated the current correlation function in the subspace of $l$th
LL and clarified the ${\bf k}$-dependence of the LL projected current
correlation function.
Then, we found that the dominant resonance of the density
correlation function occurs not at $\omega=\omega_{c}$ but at
$\omega=\omega_{{\rm NG}}({\bf k})$ at the small ${\bf k}$.
The density correlation function satisfies f-sum rule in the 
subspace of $l$th LL, then the SMA is reasonable. 
Using the current correlation function, 
we saw that the new  highly anisotropic photon energy absorption 
occurs at the NG mode frequency. The photon energy absorption rate at
$\omega=\omega_{{\rm NG}}({\bf k})$  depends on the incident
direction of the electromagnetic wave. A finite energy 
absorption occurs when the incident direction is perpendicular to the
stripes. When the incident  direction is parallel to the stripes,
the absorbed energy diverges in the $|{\bf k}|\rightarrow 0$ limit.

When the incident direction of the electromagnetic wave
is perpendicular to the stripes,
the photon energy absorption rate at
$\omega=\omega_{NG}({\bf k})$
is proportional to $B^{3/2}$ and is larger than the photon absorption
rate at $\omega=\omega_{c}$.
At  $B=2.5[{\rm T}]$ where the anisotropic resistivity at $\nu=2.5$ is 
observed, the striped Hall gas in the half-filled third LL
 absorbs the photon energy at the NG mode frequency which is twice as large
as the absorbed energy at  the cyclotron frequency.

When there is a finite excitation energy gap, the sharp absorption
only occurs at the cyclotron frequency and is not affected by
the electron interactions due to the Kohn's theorem.
Without the LL projection, we found  a sharp photon energy
absorption at $\omega=\omega_{c}$ in the striped Hall gas. 
The photon energy absorption rate at $\omega=\omega_{c}$ does not
depend on the incident direction of the electromagnetic wave in 
the $|{\bf k}|\rightarrow 0$ limit.
The cyclotron resonance is not affected in the striped Hall gas.
The result   supplements the Kohn's theorem in the system
of the NG mode.  

We used the SMA in the present paper. The SMA is one of
 two approaches to study the low energy excitation.
The other approach is 
 based on the edge picture in which
 the low energy excitations are assumed as 
small displacements of the edges of the stripes. 
 At a small ${\bf k}$, 
the excitation energy in the SMA   is proportional to 
$|\hat{k}_{y}|(A_{0}\hat{k}_{x}^{2}+B_{0}\hat{k}_{y}^{4})$ where $A_{0}$
and $B_{0}$ are constants.
 The excitation energy based on the edge picture  is proportional to 
$|k_{y}|\{(Yk_{x}^{2}+K k_{y}^{4})/|{\bf k}|\}^{1/2}$ where $Y$ and
$K$ are the compression and the bending elastic
moduli~\cite{phonon1,rev}, respectively.
The SMA excitation energy is smaller than the excitation energy
based on the edge picture.
Hence the SMA is appropriate for the study of the NG mode,
and we applied the SMA  in this paper.

In summary, 
We found that the anisotropic photon energy absorption occurs at
$\omega=\omega_{NG}({\bf k})$ in the small ${\bf k}$ region.
The photon energy absorption rate at $\omega=\omega_{NG}({\bf k})$
 depends on the incident
direction of the electromagnetic wave,
whereas the photon energy absorption rate at $\omega=\omega_{c}$
does not depend on the direction in the $|{\bf k}|\rightarrow 0$ limit. 
The absorbed energy at
$\omega=\omega_{NG}({\bf k})$ is larger than the absorbed energy
at $\omega=\omega_{c}$ when the incident direction is perpendicular to
the stripes. 
We hope that the  anisotropic photon energy absorption 
will be observed in experiments for the evidence of spontaneous breaking
of the magnetic  translational symmetry in  the striped Hall gas.

\begin{acknowledgements}
This work was partially supported by the special Grant-in-Aid for
Promotion of Education and Science in Hokkaido University and
the Grant-in-Aid for Scientific Research on Priority area 
(Dynamics of Superstrings and Field Theories)(Grant No.13135201),
provided by Ministry of Education, Culture, Sports, Science,
and Technology, Japan, and by Clark Foundation and Nukazawa Science
Foundation. One of the authors (T. A.) is supported as Institute 
Henri Poincar\'e  postdoctoral
  position, and thanks LPTMS (Orsay) for their hospitality.
\end{acknowledgements}

\appendix
\section{ Definition of a photon energy absorption rate}
In this section, 
we derive a photon energy absorption rate of the QHS
 by using the first order
perturbation  when an electromagnetic wave is added.
It is assumed that the electromagnetic wave 
${\bf E}={\bf \cal E}e^{i{\bf k}_{3D}\cdot{\bf r}_{3D}-i\omega t}$ 
is propagating  with the wave number vector
${\bf k}_{3D}=({\bf k},k_{z})$ and oscillating with the frequency
$\omega$ in the spacetime $({\bf r}_{3D},t)=({\bf r},z,t)$. 
Then the interaction
Hamiltonian 
$V_{int}(t)=e\int d^{3}r {\bf A}_{ext}({\bf r}_{3D})\cdot
{\bf j}_{3D}({\bf r}_{3D})$
is written with ${\bf A}_{ext}={\bf \cal E}
                e^{i{\bf k}_{3D}\cdot{\bf r}_{3D}-i\omega t}/i\omega$
as      $V_{int}(t)=e\int d^{3}r {\bf \cal E}\cdot{\bf j}_{3D}({\bf r}_{3D})
                  e^{i{\bf k}_{3D}\cdot{\bf r}_{3D}-i\omega t}/i\omega
     =e{\bf \cal E}\cdot\tilde{{\bf j}}_{3D}({\bf k}_{3D})
e^{-i\omega t}/i\omega$.
Here we use a three-dimensional current operator
${\bf j}_{3D}=
(j^{1}({\bf r}_{3D}),j^{2}({\bf r}_{3D}),j^{3}({\bf r}_{3D}))$ with 
$j^{\mu}({\bf r}_{3D})=j^{\mu}({\bf r})\delta(z)$ for $\mu=0,1,2$,
and $j^{3}({\bf r}_{3D})=0$.
 Then the Fourier transformation gives 
$\tilde{j}^{\mu}({\bf k}_{3D})=\tilde{j}^{\mu}({\bf k})$ and
$\tilde{j}^{3}({\bf k}_{3D})=0$. 
By summing up all possible final states, a transition probability
from an initial state $|i\rangle$ to a final state $|f\rangle$ 
is given by using the first order perturbation  as follows,
\begin{eqnarray}
Prob&=&\sum_{f}\left|-i
\int_{-T/2}^{T/2}dt\langle f|V_{int}(t)|i\rangle e^{i(E_{f}-E_{i})t}\right|^{2}
\nonumber\\
&=&e^{2}\sum_{l,m=1,2}\frac{{\cal E}_{l}{\cal E}_{m}}{\omega^{2}}
   \sum_{f}F_{T}(\omega)^{2}
\langle i|
\tilde{j}^{l}(-{\bf k})|f\rangle\langle f|
\tilde{j}^{m}({\bf k})|i\rangle.
\end{eqnarray}
Here 
$F_{T}(\omega)=\int_{-T/2}^{T/2}dt e^{-i(\omega-E_{f}+E_{i})t}$,
which satisfies
$\lim_{T\rightarrow
\infty}F_{T}(\omega)=2\pi\delta(\omega-E_{f}+E_{i})$.
The transition probability per unit time is written as
\begin{eqnarray}
\frac{Prob}{T}
&=&  \sum_{l,m=1,2} \sum_{f}2\pi\delta(\omega-E_{f}+E_{i})
\frac{e^{2}{\cal E}_{l}^{\ast}{\cal E}_{m}}{\omega^{2}}
\langle i|
\tilde{j}^{l}(-{\bf k})|f\rangle\langle f|
\tilde{j}^{m}({\bf k})|i\rangle.
\label{eqn:tp}
\end{eqnarray}
Suppose that the initial state $|i\rangle$ is  the ground 
state $|0\rangle$.
Then  Eq.~(\ref{eqn:tp}) is represented by the current correlation
function as 
\begin{eqnarray}
\frac{Prob}{T}
&=&\sum_{l,m=1,2}
\frac{2e^{2}{\cal E}_{l}^{\ast}{\cal E}_{m}{\cal A}}{\omega^{2}}
{\rm Im}_{\omega>0}
   \tilde{\Pi}_{0}^{lm}({\bf k},\omega).
\end{eqnarray}
We define the photon energy absorption rate  $P$ as
\begin{eqnarray}  
\omega \frac{Prob}{T\cdot {\cal A}}
=\sum_{l,m=1,2}\frac{2e^{2}{\cal E}_{l}^{\ast}{\cal E}_{m}}{\omega }
{\rm Im}_{\omega>0}\tilde{\Pi}_{0}^{lm}({\bf k},\omega).
\end{eqnarray}

Suppose that the
electromagnetic wave is the left-handed circularly polarized 
wave with ${\bf \cal E}={\cal E}_{L}({\bf e}_{x}+i{\bf e}_{y})$, 
then the photon energy absorption rate
is given by
\begin{eqnarray}
P=\frac{2e^{2}{\cal E}_{L}^{2}}{\omega }
\bigl({\rm Im}_{\omega>0}\tilde{\Pi}_{0}^{xx}({\bf k},\omega)
+{\rm Im}_{\omega>0}\tilde{\Pi}_{0}^{yy}({\bf k},\omega)\bigr).
\label{eqn:p}
\end{eqnarray}

\section{ Cyclotron resonance}
In this section, we derive the cyclotron resonance when a homogeneous
electromagnetic wave is added to the striped Hall gas.
The covariant momentum  is written as
 \begin{eqnarray}
\pi^{i}&=&m\tilde{j}^{i}({\bf 0})
\end{eqnarray}
with 
$\tilde{j}^{i}({\bf k})=\int d^{2}r\Psi^{\dag}({\bf r})v^{i}\Psi({\bf r})
e^{i{\bf k}\cdot{\bf r}}$,
where $i=x,y$ and  $v^{i}=\omega_{c}(-\eta,\xi)$.
   The covariant momentum operators satisfy 
   $[\pi^{x},\pi^{y}]=-i2\pi Q=-i2\pi N_{e}$,
   where we replace the conserved charge $Q$ by $N_{e}$.
   The covariant momentum operators also satisfy the following relation 
   $i[H,\pi^{i}]=-2\pi \sum_{j=x,y}\epsilon_{ij}\pi^{j}/m$,
   where $\epsilon_{ij}$ is antisymmetric tensor and suffix $i,j$ run 
   $x,y$.
   From these relations,  we now define 
    the energy ladder operators $\pi_{\pm}$ as
   \begin{eqnarray}
   \label{eqn:energypmdif}
   \pi_{\pm}&=&\frac{1}{\sqrt{4\pi N_{e}}}(\pi^{x}\pm i\pi^{y}).
   \end{eqnarray}
   The energy ladder operators satisfy 
   $[\pi_{-},\pi_{+}]=1$.
   Then we find that
   \begin{equation}
   [H,\pi _{\pm}]=\pm\omega _{c} \pi _{\pm}.
   \label{eqn:energypm}
   \end{equation}
Thus, $\pi_{\pm}$ shifts the energy of the system by the cyclotron energy  
   $\omega _{c}$.
This property indicates that an excited state with the  energy
    $E_{0}+n\omega_{c}$ exists, where $H|0\rangle=E_{0}|0\rangle$ and $n$ is 
   an integer. 
   In the following discussion, we assume that 
   $\pi_{-}|0\rangle =0$ and
   $\langle 0|\pi_{+}=0$,   
   because of the stability of the ground state. This assumption prohibits
   the transfer to lower Landau levels than the ground state.

Next, we calculate the photon energy absorption rate in 
Eq.~(\ref{eqn:p}) with ${\bf k}=0$.
By using  Eq.~(\ref{eqn:correlation2}),
${\rm Im}_{\omega>0}\tilde{\Pi}_{0}^{ij}({\bf 0},\omega)$ is written as
\begin{eqnarray}
\sum_{\sigma}
4{\cal A}\int\frac{d^{2}k^{\prime}}{(2\pi)^{2}}
   \frac{d^{2}k^{\prime\prime}}{(2\pi)^{2}}
\frac{\sin{(\hat{k}_{x}^{\prime}/2)}}{\hat{k}_{x}^{\prime}}
\frac{\sin{(\hat{k}_{x}^{\prime\prime}/2)}}{\hat{k}_{x}^{\prime\prime}}
\langle {\bf 0},\sigma_{0}|\tilde{j}^{i}({\bf k}^{\prime})|
{\bf 0},\sigma\rangle\langle {\bf 0},\sigma|\tilde{j}^{j}
({\bf k}^{\prime\prime})|{\bf 0},\sigma_{0}\rangle
\pi\delta(\omega-E_{\sigma}({\bf 0})-E_{\sigma_{0}}({\bf 0})).
\label{eqn:picyclo}
\end{eqnarray}
We evaluate $\langle {\bf 0},\sigma_{0}|\tilde{j}^{i}({\bf k}^{\prime})|
{\bf 0},\sigma\rangle$
where $|{\bf 0},\tau\rangle=\pi_{+}|{\bf 0},\tau_{0}\rangle$
and $|{\bf 0},\sigma_{0}\rangle$
is  the ground state in the HFA.
The current operator is written in the vNL basis as 
\begin{eqnarray}
\tilde{j}^{i}({\bf k})
&=&\int_{\rm BZ}\frac{d^2 q}{(2\pi)^{2}}
\sum_{l,l^{\prime}}
b_{l}^{\dag}({\bf q})b_{l^{\prime}}({\bf q}+\hat{\bf k})
\langle f_{l}|\frac{1}{2}\{v^{i},e^{-ik\cdot\xi}\}
|f_{l^{\prime}}\rangle e^{\frac{i\hat{k}_{x}}{4\pi}(2q_{y}+\hat{k}_{y})}.
\end{eqnarray}
Taking into account the forbidden transition between occupied states
or between empty states,
the matrix element in  Eq.~(\ref{eqn:picyclo}) is given by 
\begin{eqnarray}
\langle {\bf 0},\sigma_{0}|\tilde{j}^{i}({\bf k}^{\prime})|
{\bf 0},\sigma\rangle
=\frac{1}{m{\cal A}}\langle {\bf 0},\sigma_{0}|\pi^{i}|{\bf 0},\sigma\rangle
(2\pi)^{2}\delta({\bf k}^{\prime})+ \sum_{{\bf n}\neq{\bf 0}}
a_{{\bf n},\sigma}^{i}
(2\pi)^{2}\delta({\bf k}^{\prime}+2\pi{\bf n}).
\end{eqnarray}
Inserting the matrix element into  Eq.~(\ref{eqn:picyclo}),
we obtain  
\begin{eqnarray}
\sum_{i=x,y}{\rm Im}_{\omega>0}\tilde{\Pi}^{ii}_{0}({\bf 0},\omega)
&=& \frac{1}{{\cal A} m^{2}}\sum_{i=x,y}\sum_{\sigma}
\langle {\bf 0},\sigma_{0}|\pi^{i}|{\bf 0},\sigma\rangle
\langle {\bf 0},\sigma|\pi^{i}|{\bf 0},\sigma_{0}\rangle
\pi\delta(\omega-E_{\sigma}({\bf 0})-E_{\sigma_{0}}({\bf 0}))
=\frac{2\pi\rho_{e}}{m^{2}}\pi\delta(\omega-\omega_{c}), 
\end{eqnarray}
where $\rho_{e}=N_{e}/{\cal A}$.
 $a_{{\bf n},\sigma}^{i}({\bf n}\neq {\bf 0})$ does not contribute to
the current correlation function.
By inserting this current correlation function into
 Eq.~(\ref{eqn:p}) with ${\bf k}={\bf 0}$,
a sharp energy absorption is obtained as follows,
\begin{eqnarray}
P_{\rm cyclotron}=\frac{e^{2}{\cal E}_{L}^{2}\rho_{e}}{m}
2\pi\delta(\omega-\omega_{c}).
\end{eqnarray} 



\end{document}